\documentclass[11pt]{article} 

\usepackage{floatrow} 
\usepackage[utf8]{inputenc} 

\usepackage{geometry} 
\usepackage{graphicx} 

\usepackage{xcolor}
\usepackage{authblk}

\usepackage{graphicx}
\usepackage{amsfonts}
\usepackage{textgreek}
\usepackage{booktabs} 
\usepackage{array} 
\usepackage{paralist} 
\usepackage{verbatim} 
\usepackage{subfig} 

\usepackage{fancyhdr} 
\usepackage{sectsty}
\allsectionsfont{\sffamily\mdseries\upshape} 

\usepackage[nottoc,notlof,notlot]{tocbibind} 
\usepackage[titles,subfigure]{tocloft} 

\title{Milestones on the Quantum Utility Highway}
\author{Catherine C. McGeoch and Pau Farr\'{e} }
\affil{D-Wave Systems Inc.} 
\affil{ \em \{cmcgeoch, pfarre\}@dwavesys.com} 

\begin{document}
\maketitle


\begin{abstract}

We introduce {\em quantum utility}, a new approach to evaluating quantum performance  
that aims to capture the user experience by including overhead costs associated with the quantum 
computation.   A demonstration of quantum utility by a quantum processing unit (QPU) shows that the QPU can outperform classical solvers at some tasks of interest to practitioners, when considering computational overheads.  We consider overhead costs that arise in standalone use of the QPU (as opposed to a hybrid computation context).  We define three early milestones on the path to broad-scale quantum utility that focus on restricted subsets of overheads:  Milestone 0 considers pure anneal time (no overheads) and has been demonstrated in previous work;  Milestone 1 includes overhead times to access the QPU  (that is, programming and readout); and Milestone 2 incorporates an indirect cost associated with minor embedding. 

We evaluate the performance of a D-Wave Advantage QPU with respect to Milestones 1 and 2, using a testbed of 13 input classes and seven classical solvers implemented on CPUs and GPUs.  For Milestone 1,  the  QPU outperformed all classical solvers in 99\% of our tests.  For Milestone 2, the  QPU outperformed all classical solvers in 19\%  of our tests, and the scenarios in which the QPU found success correspond to cases where classical solvers most frequently failed.    

Analysis of test results on specific inputs reveals fundamentally distinct underlying mechanisms that 
explain the observed differences in quantum and classical performance profiles.  We present evidence-based arguments that these distinctions bode well for future annealing quantum processors to support 
demonstrations of quantum utility on ever-expanding classes of inputs and for more challenging milestones. 
\end{abstract} 

\section{Introduction} 
Annealing-based QPUs manufactured by D-Wave\footnote{D-Wave trademarks and registered trademarks used herein include D-Wave, Leap, Ocean, Chimera, Pegasus, Zephyr,  Advantage,  D-Wave Two, D-Wave 2X, D-Wave 2000Q, and Advantage2.  Intel and Xeon are trademarks of Intel Corporation.  NVIDIA and NVIDIA GeForce GRX are trademarks of NVIDIA Corporation.}  are designed to heuristically solve problems in combinatorial optimization.  
These systems operate within a declarative problem-solving paradigm: rather than implementing a program to solve a given problem,  the user reformulates the problem to match a solver already implemented in quantum hardware and firmware.   Reformulation creates subtasks, including translating the original input $X$ into a {\em logical} graph input $G$ that  matches the problem solved natively, and minor-embedding\footnote{In graph theory, an embedding maps a graph $G$ to another graph $P$, and a minor embedding maps $G$ onto a minor of $P$ (including $P$).  Here we use both terms interchangeably to refer to the latter task.} $G$ to create a {\em physical} input $P$ that matches the qubit connectivity structure in the QPU.  

\begin{figure}
\begin{centering}
\includegraphics[width=\textwidth]{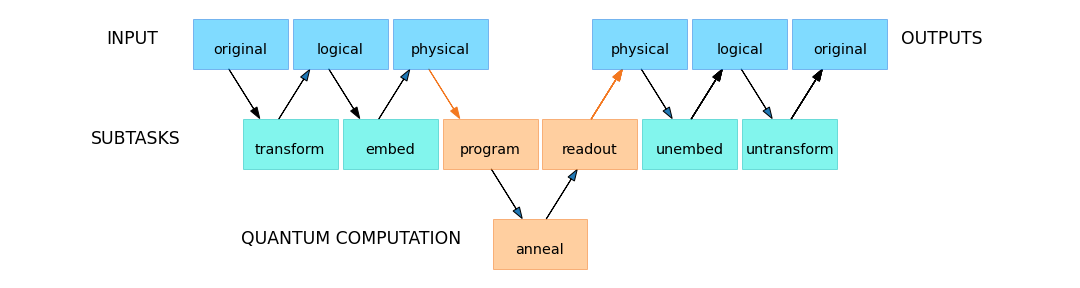} \\
\end{centering}
\caption{Subtasks associated with using a quantum annealing system. Input and output formulations are marked in blue;  subtasks in teal are associated with problem translation.   Tasks in orange correspond to access time, which is considered in Milestone 1.  The orange arrows identify network transmission overheads in a cloud-based service model.} 
\label{fig::layers} 
\end{figure} 

Figure \ref{fig::layers} illustrates these subtasks,  which create overhead costs that increase total computation time;  the purely quantum component of this workflow is called an anneal.  Note that this diagram applies only to a sequential workflow that arises in standalone use of the QPU.  We do 
not consider hybrid computations, in which quantum and classical components operate concurrently on distributed platforms,  creating different workflows and cost breakdowns.   

We introduce the concept of {\bf quantum utility}, which aims to capture quantum performance 
as experienced by the user.  A demonstration of quantum utility is an affirmative answer to this 
question: {\em Considering overheads, can the quantum system outperform classical alternatives at some task of application interest?}  Quantum utility is demonstrated when the quantum computation is fast enough to compensate for classical overhead costs of using the quantum system.   

A demonstration of quantum utility over a broad range of overheads and inputs must wait upon some future date (see Section \ref{conclusions}).   In the meantime, we identify milestones along the way:  a milestone may be thought of as a test of quantum utility in a restricted context associated with specific overheads.  We define the first three milestones as follows. 
\begin{enumerate}
\item[] {\bf Milestone 0, no overheads:} A demonstration that the QPU outperforms classical solvers 
that read the same (physical) inputs, measuring quantum annealing time.  Section \ref{m0} reviews previous publications containing demonstrations of Milestone 0.  We do not explicitly address it here, except to point out that a demonstration of Milestone 1 or 2 implies a demonstration of Milestone 0.    

\item[] {\bf Milestone 1, access time:}  A demonstration that the QPU outperforms classical solvers reading the same
(physical) inputs, when measuring QPU access time.  Access time includes a {\em programming} step that occurs at least once per input, an {\em anneal} step that occurs once per output, and a {\em read} step that occurs once per output. On the Advantage system tested here, programming imposes a lower bound access cost of approximately 16 ms, before the first anneal can begin.  

\item[] {\bf Milestone 2, indirect cost of embedding:}  A demonstration that the QPU, which reads a larger physical input, 
can outperform a classical solver that reads a smaller logical input, when measuring access time.  That is, the minor embedding step maps a {logical input} defined on an arbitrary graph $G$ (with $n$ nodes and $m$ edges), onto a {physical input} defined on a fixed hardware graph $P$ (with $q$ qubits and $c$ couplers), such that $n+m \leq q + c$.  This incurs 
both a direct cost to compute the mapping (not considered here),  and an indirect cost in the sense that classical 
solvers may exploit a relative speedup due to this size differential.  
\end{enumerate}  

Milestones 1 and 2 are arguably most important to achieving the above-mentioned 
consensus on quantum utility,  because the relevant overheads are unique to quantum annealing and are common to most users' experience. That is, depending on the envisioned use case, the other overhead times in Figure \ref{fig::layers} may be considered negligible or otherwise irrelevant to a performance comparison.  
Section \ref{scope} describes some application scenarios for which Milestones 1 and 2 might be considered sufficient to demonstrate real-world quantum utility.   

Section \ref{setup} describes our test setup.  We evaluate performance of an Advantage QPU with respect to 
Milestones 1 and 2, in comparison to seven classical solvers: two read physical inputs $P$ and five read general 
inputs $G$;  three run on GPUs and four run on CPUs. Our tests use 25 inputs each from 13 different classes.  
For each instance, solvers must return a sample of $s \in \{1,10,100,1000\}$ independent solutions within a time limit of $t \in \{ .02, .05, .1,.2,  .5, 1\}$ seconds (that is, allowing up to $t/s$ seconds per solution, depending on initialization overheads).  A test {\em scenario} corresponds to a specific combination of $s$ and $t$.   

We perform 247 tests combining 19 scenarios and 13 input classes, and rank solvers according to the quality of solution samples returned.  For each scenario and input class, a {\em win} is awarded to a solver that finds solutions strictly better than the others.  A {\em fail} is awarded if the solver is unable to return a complete set of $s$ solutions within time limit $t$.  
Note that a few tests had multiple winners, and it is possible for a solver to neither win nor fail a given test.   Results are summarized below. 
\begin{itemize}
\item Section \ref{resultsm1} presents Milestone 1 results. The Advantage QPU wins in 99\% of tests (that is, all 
but 3 of 247).  Some classical solvers fail in scenarios where $t/s$ is smallest.   

\item Section \ref{resultsm2} presents Milestone 2 results for eight input classes that have both pre-embedded ($G$) 
and post-embedded ($P$) versions.  We compare the QPU, reading $P$, to five classical solvers, reading $G$, for 
a total of 152 tests in 19 scenarios.  The QPU wins in 19\% of the tests.  The tests where Advantage wins 
correspond to largest $n$ and smallest $t/s$; these are the same tests for which classical solvers most frequently fail.  

\item Section \ref{discussion} identifies some fundamental differences in the computational mechanisms that drive quantum and classical performance,  which explain the observed contrary trends in quantum versus classical patterns of wins and losses.  We present evidence-based arguments that these fundamental differences bode well for demonstrations of these two milestones on more varied types of inputs, 
as well as demonstrations of more challenging milestones, using future annealing-based QPUs.    
\end{itemize}

We remark that these results represent a watershed moment in the development of quantum annealing processor technologies over the past decade.  With the few exceptions mentioned in Section \ref{previous},  previous-generation QPUs were too small, and access times too high, to support positive outcomes on milestone tests such as these.  As discussed in Section \ref{inputs}, Advantage QPUs have reached a threshold of size and connectivity sufficient to hold inputs that require more than 16 ms to be solved classically;  as a result,  Milestones 1 and 2 (and beyond) have become feasible to test.     

Nevertheless,  these are just the first steps of a longer journey toward routine demonstrations of quantum utility over a broad range of overheads and inputs.  As discussed in Section \ref{inputs},  our tests using a current-generation Advantage QPU are incomplete because other input classes can be identified for which logical inputs remain too small to serve as viable test candidates.   See Section \ref{conclusions}  for more about this point.  

\begin{figure}
\begin{centering}
\begin{tabular}{|lcc|lll|} \hline 
{\bf Year} & {\bf Source}  & {\bf QPU}  & {\bf Milestone} & {\bf Metric}  & {\bf Solvers}  \\  \hline  \hline
2013 & \cite{mcgwan}  & D-Wave Two  & 0$^*$,1     & A      & EX(1)  HS (1) PS(1)      \\  
2015 & \cite{trukoc}    &  D-Wave 2X   & 0                 & B      &  EX(2) HS(3)   \\  
2015 & \cite{kingttt}         & D-Wave 2X  &  0, 1         & A  &  HS(2)    \\   
2016 & \cite{denchev}  & D-Wave 2X      & 0, $1^*$   & C        &  HS(1) MC(1) \\    
2017 & \cite{mcclauses}  &  D-Wave 2000Q        &  0, 1          &  B       &  HS(1)       \\  
2019 & \cite{jking19}     &  D-Wave 2000Q         & 0, 1          & C \& D &  HS(2) MC (2)  \\  
2019 & \cite{koshka}     & D-Wave 2000Q          &  0,1                   &  D  & HS(1) \\      
2019 & \cite{pang}        & D-Wave 2000Q          & 0$^*$, 1      & B         &  EX(3) HS(2)  MC(1)  \\ 
2020 & \cite{jones21}    &  D-Wave 2X  & 0                & B    &  EX(1) HS(1)     \\ 
2020 & \cite{inoue20}    & D-Wave 2000Q        &  0  &   B  &  PS(2) \\ 
2021 & \cite{kingsim}    & D-Wave 2000Q        & 0, 1$^*$            &E          &  MC(1)    \\   
2021 & \cite{abel2021}  & Advantage  & 0              & C         &  HS(3) \\ 
2022 & \cite{raisuddin}  &  D-Wave 2000Q      &  0           & C           & HS(1)   \\  
2022 & \cite{tasseff22}  & (several)     & $0,1$    &  C          & EX(1) HS(2) MC(1)   \\ 
2022 & \cite{liu22}        &  D-Wave 2000Q      &  0            &  B        & PS(1) \\  \hline 
\end{tabular}\\
\end{centering} 
\caption{Previous work demonstrating Milestones 0 and 1 using past- and current-generation D-Wave QPUs.  
An asterisk marks a milestone that was not explicitly addressed but can be inferred from the published results.  
The Metric column refers to the figure of merit used in the evaluation, and the Solvers column 
lists categories of classical solvers used for comparisons.  Column entries are decoded in the text. } 
\label{fig::m0} 
\end{figure} 

\subsection{Previous Work} 
\label{previous} 
\label{m0}

Figure \ref{fig::m0} lists published papers that meet our criteria for  
demonstrating Milestone 0 (and in some cases Milestone 1):  they measure {both} solution 
quality and computation time,  and report superior QPU performance on a significant proportion of inputs tested.  These results span a variety of application domains (including quantum simulation, combinatorial optimization, and diverse sampling),  input types, performance metrics, and classical competition:  we consider such variety to be necessary to reaching consensus 
about whether quantum utility has been demonstrated. 

The {QPU} column refers to previous and current generation D-Wave QPUs, with nominal qubit counts as follows:  D-Wave Two (512), D-Wave 2X (1000+),  D-Wave 2000Q  (2000+), and Advantage (5000+).  The first three generations were based on the Chimera connection topology, with $6$ couplers per qubit;  the Advantage generation of QPUs is based 
on the Pegasus topology with 15 couplers per qubit.

In the {Milestone} column, an asterisk marks a milestone that was not explicitly 
measured in the paper but can be inferred from the results.  For example, Milestone 1 implies 
Milestone 0 because anneal time is always less than access time.  Milestone 1 can be inferred 
from Milestone 0 when a reported computational speedup is large enough (that is, 10,000-fold or higher) to easily account for access overhead times.   
Tasseff et al. \cite{tasseff22} test on several available D-Wave 2000Q and Advantage systems; they 
consider Milestones 0 and 1, as well as overhead costs of network transmission and queuing time, which constitutes a milestone somewhere beyond Milestone 2. (Our Milestone 2 is not relevant to their work because inputs are not minor embedded.)   

The {Metric} column identifies the type of performance metric(s) used in each study, as follows. 
\begin{enumerate}\itemsep-.4em
\item[A] Best solutions returned within a fixed time limit 
\item[B] Better solutions faster, considering both solution quality and runtime
\item[C] Fastest time to an optimal solution 
\item[D] Fastest time to sample all near-optimal valleys in the solution space 
\item[E] Fastest convergence of solution samples to a target distribution 
\end{enumerate} 

The {Solvers} column identifies categories of classical solvers tested in each work; numbers in parentheses refer to 
multiple versions of solvers within the category, as follows. 
\begin{itemize}
\item Exact (EX) solvers guarantee to return optimal solutions if given enough computation time.  When used in tests 
of heuristic performance that accept non-optimal solutions (as with metrics A, B, D, E), these solvers report working 
solutions found at earlier stopping times.  Algorithms of this type include Branch-and-Bound, Integer Linear Programing, and Message Passing. 

\item  Heuristic Search (HS) solvers typically start with a random candidate 
solution and then iterate, making incremental changes to the candidate while seeking a downhill path 
towards better-quality solutions.  HS solvers vary according to, e.g., mechanisms for choosing next-moves,  strategies for escaping local minima, or the number of multiple solution paths being pursued.  HS solvers in the published works include Greedy,  the Hamze-de Frietas-Selby (HFS) method \cite{hamzede, selbysrc}, Simulated Annealing, TABU search,  Parallel Tempering, and population-based (genetic) methods.  

\item Monte Carlo {\bf (MC)} search represents an important subclass of HS methods.  
An MC solver performs a guided random walk to explore the solution landscape using a 
probabilistic strategy that simulates a natural process.  Variations of MC in the published works of Figure \ref{fig::m0} include Markov Chain MC, path integral MC, quantum MC, spin vector MC,  and Wolff cluster MC.  (The Simulated Annealing and Parallel Tempering heuristics arguably 
belong to this paradigm, although implemented solvers can vary widely in their fidelity to a natural model; our classification follows authors' nomenclatures.)  

\item Problem Specific {\bf (PS)} solvers read  original inputs $X$ for a specific problem domain, rather than translated logical $G$ or physical $P$ graphs.  The differential in problem size between $X$ and $P$, together with measurement of access time, would correspond to a milestone somewhere beyond our Milestone 2.  
\end{itemize} 

\paragraph{Other empirical work.} 
A number of  papers that compare performance of D-Wave QPUs against classical solvers are not included in Figure \ref{fig::m0}, for reasons briefly summarized below. 

\begin{itemize} 
\item Several papers consider a performance metric known as {\em quantum speedup} \cite{albashspeedup, hen15, katz14,katz15,mandra18, roennow, zhu16}.  This body of work aims to evaluate how computation time {\em scales} with problem size over a finite range.  
Scaling analyses look at the shape of a curve showing computation time versus problem size, 
but reported times are not intended to accurately reflect total computation times.  Instead, 
these papers typically report times for a core operation (the part that scales with $n$), 
and some evaluate classical performance on a hypothetical parallel platform (dividing core times by $n$).  Furthermore, these papers look at QPU performance against the ``best available''  classical solver, which involves extensive precomputation to optimize solver parameters, contrary to the Fair Test policy described in Section \ref{params}.  Lacking information 
about true computation times under realistic use scenarios, we cannot determine 
whether Milestone 0 has been demonstrated in these papers. 

\item Some papers compare quantum and classical solution quality  --- reporting both positive and negative results --- but omit computation times;  others report computation times but apply unequal levels of computational work
\cite{higham22, inoue21, juenger, kim20, yarkoni21}.  Some negative results in these papers 
can be attributed to small QPU size, as discussed in the introduction.  In any case, without a way to calibrate 
computational work, we cannot determine which solvers would give superior results under comparable 
use conditions;  see Section \ref{discussion} for discussion of this issue.  
\end{itemize}  

\subsection{Overhead Costs of QA}
\label{overhead} 

This section describes two categories of overhead costs that are considered in our benchmark tests.  

\begin{figure}
\begin{centering} 
\includegraphics[width=\textwidth]{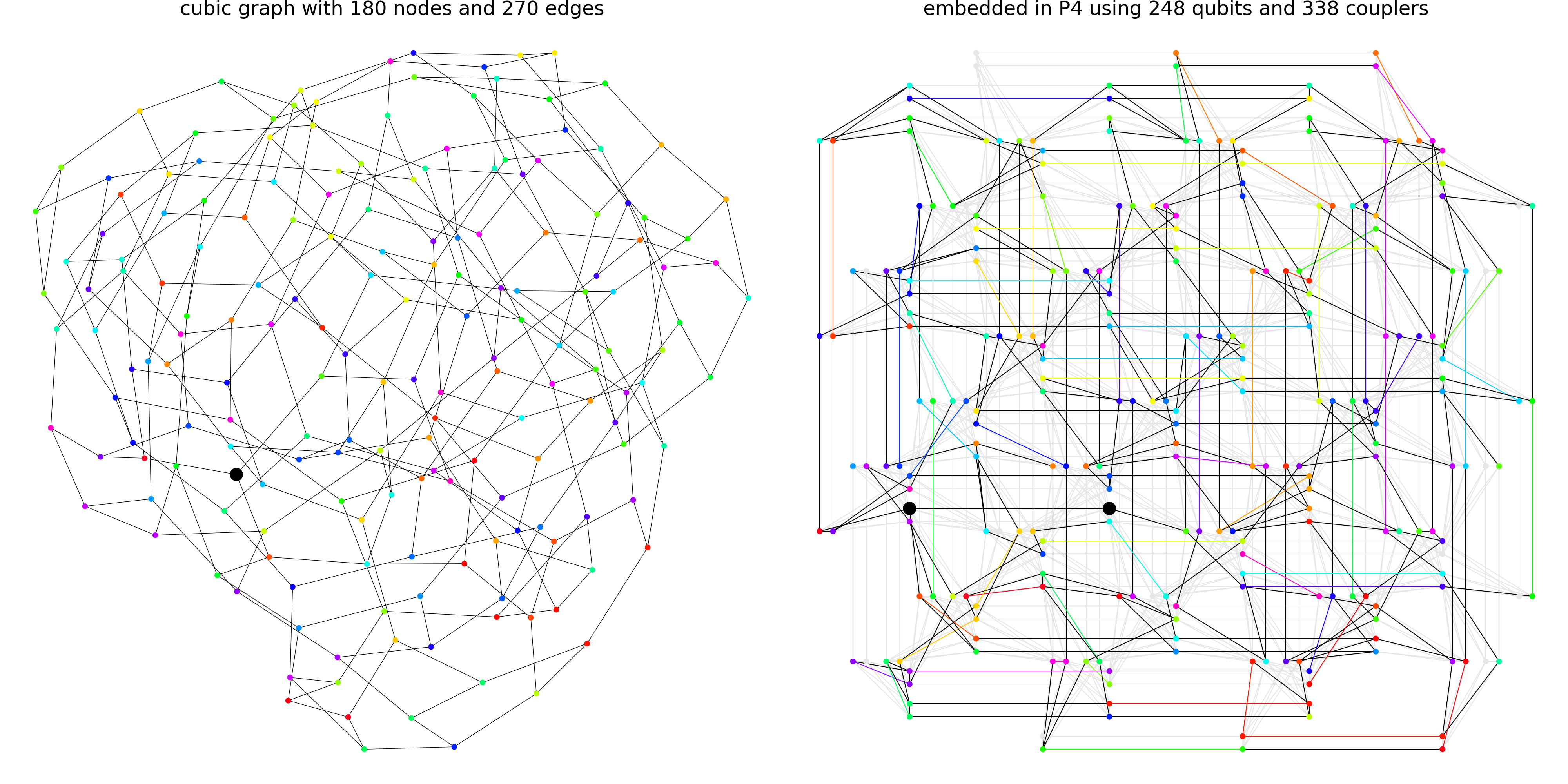}\\
{ (a) \hspace{3in} (b) } \\
\end{centering} 
\caption{Minor embedding.  (a) General graph $G$.  (b) A minor embedding of $G$ onto a Pegasus graph $P$.  
The highlighted black node in $G$ is split into a chain of two nodes in $P$;  both the original node and the chain are connected to three neighbors (pink, green, red). Each qubit in Pegasus is connected by couplers to 15 
neighbors (fewer on boundaries). Qubits and couplers that are unused in the embedding are shown in pale gray. }
\label{fig::embed}  
\end{figure} 

\subsubsection{Minor Embedding} 
\label{embedov} 
Milestone 2 considers an indirect cost associated with minor embedding as follows. 
The Pegasus graph defines the qubit and coupler connectivity structure inside all Advantage QPUs.  The specific {hardware graph} inside any given QPU is a subgraph of Pegasus, due to a small proportion of disabled qubits and couplers that do not meet specifications; see Section \ref{solvers} for details about the system used in our tests. 

Given an arbitrary logical graph $G$, minor embedding maps each node of $G$ to a {\em chain} of one or more connected qubits in the hardware graph. (Despite the name, they need not be connected in a strictly linear sequence, and tree-like chain structures are common.)  Minor embedding is necessary for any $G$ that is not a subgraph of the hardware graph; in particular, minor embedding is useful for mapping high-degree nodes of $G$ onto limited-degree nodes of $P$.   For example, a logical node $x$ of degree 20 can map to a two-node chain $(q1,q2)$, each of degree 10, plus an additional chain edge $(q1,q2)$ tying them together.  The {\em chain strength} $J_{chain}$ is a  weight applied to all new chain edges that are introduced by minor embedding.  Our tests use a default chain strength that is recommended by an Ocean system utility (more discussion below). 

In Figure \ref{fig::embed}, panel (a) shows a graph $G$ with $n=180$ nodes and $m=270$
edges, totaling 450 components.  Panel (b) shows $G$ after being minor-embedded onto a Pegasus 
graph containing $q=248$ qubits and $c=338$ couplers, totaling 586 components.  
Each node in $G$ is mapped to an equal-colored chain of one or more nodes in $P$; for 
example,  the oversized black node with three neighbors (green, orange, red) 
is represented by a two-qubit chain in the embedding.  In terms of total {\em input size} 
(nodes plus edges),  the embedded graph has $586=248+338$ qubits and couplers, about 1.4 times larger than $G$. 

Minor embedding incurs two types of overhead costs:  (a) the computation time to find a mapping 
from $G$ to $P$ and apply it; and postprocessing time to unmap solutions for $P$ back to solutions 
for $G$ (a direct cost 
that is not considered in this paper), and (b) the indirect cost considered in Milestone 2, whereby classical solvers reading $G$ can work on a smaller input of size $(n,m)$, while the QPU works on a larger input of size $(q,c)$.  Under the reasonable  
assumption that smaller inputs are easier to solve,  this size difference creates a potential runtime advantage 
for classical solvers relative to the QPU.  Our tests measure this difference by running all solvers for equal time 
limits,  and comparing solution quality with respect to the logical problem $G$,  after QPU solutions to $P$ been untranslated back to solutions for $G$.   

\subsubsection{Access Time}
\label{accessov}  
Milestone 1 incorporates {\em access time}, which, in addition to purely quantum annealing time, 
includes the time needed to transfer information on and off the quantum chip, as follows. 

First, when given a physical input graph $P$ defined by weight vectors $(h,J, J_{chain})$, the QPU performs at least one {\em programming}\footnote{Here, {\em programming} and {\em readout} are used as umbrella terms that include several subtasks. When applicable, subtask  times are set to constant default values throughout, such as  {\tt qpu\_delay\_time\_per\_sample} = $20\; \mu$s, which is performed during readouts.} step to map $(h,J, J_{chain})$ onto the analog control system that drives qubit states 
during the anneal.  On the {Advantage\_system4.1} QPU used in our tests, this operation takes 
$t_{program} = 16$ ms.   
Multiple programmings can be used to make small adjustments to the physical problem representation.  In our tests,  reprogramming was applied for two reasons:
\begin{itemize}
\item For native inputs, each programming step applies a random spin reversal transform (SRT). An SRT 
assigns the physical output states (up,down) of a subset of qubits to be either $(-1,+1)$ or $(+1,-1)$. 
This random assignment counteracts certain types of physical biases in the quantum hardware. 

\item For embedded inputs, each programming step increments or decrements chain strength within a small range of values $J_{chain} \in [0.5x, \ldots, 2x]$, where $x$ is the default chain strength suggested by an Ocean system utility.   
\end{itemize} 
In both cases, the modification can shift the distribution of sampled solution energies 
higher or lower by a small amount.  Multiple programmings have the effect of widening the range of energies in the 
full set of sampled solutions,  which may potentially move sample minimums closer to ground states.  

Second, after one anneal produces a solution to input $P$, a {\em readout} operation is performed to move the solution off the chip.  
On the Advantage system used in our work, $t_{read} \leq 0.241$ms. Smaller readout times can be observed when only a portion of 
the chip is occupied by input weights, but our tests focus on large inputs and readout times were fairly consistent. (Both programming and readout times include other low-level operations not discussed here.)  

Three annealing control parameters were varied in our tests:  the number of 
programmings $p$, anneal time per solution $t_{anneal}$, and the number of anneal-plus-readout operations $r$.  This works out to $\lfloor r/p \rfloor$ solutions per programming, with the last batch rounded up if necessary to meet a given time limit $t$.  Access time to return $r$ solution to an input instance $P$ is therefore measured as  
\begin{eqnarray}
  t_{access} & = &  p \cdot t_{prog}  +  r \cdot (t_{anneal}  + t_{read}).   
\end{eqnarray}  

\section{Experimental Setup} 
\label{setup} 

This section presents an overview of inputs, solvers,  and performance metrics used in our tests. 

\paragraph{The Problem}
The Ising Model (IM) problem is defined as follows:  given a graph $G = (V, E)$ on $n$ nodes and $m$ edges, 
together with real-valued  
weights $h = \{ h_i \}$ (called fields) on nodes and $J =\{  J_{ij} \}$ (called couplings) on edges,  assign 
spin values $ x_i \in \{-1,+1\} $ to nodes so as to minimize the energy function 
\begin{eqnarray}
 F(x)  &  = &  \sum_{i \in V}   h_i x_i   +   \sum_{(i,j) \in E}  J_{ij} x_i  x_j. 
\end{eqnarray} 

This problem is NP-hard when $G$ is nonplanar \cite{istrail}.   In physics applications that model natural
phenomena, an optimal solution is called a ground state and non-optimal solutions are called excited states. 
The Quadratic Unconstrained Binary Optimization (QUBO) problem, more familiar to researchers in 
computer science and operations research, is to optimize the same objective function (2),  
defined on {binary variables} $b_i \in \{0, 1\}$ instead of spins.  Under the trivial 
transformation $x_i = 2b_i -1$, the objective functions for QUBO and IM have equivalent cost spectra, except for a constant 
offset that varies by instance.  

Either formulation can be used as input to the QPU, and we generically refer to both as binary quadratic models (BQMs).  Our input generators construct BQMs of either type, according to their motivating application domains. 

\subsection{Inputs} 
\label{selected} 
\label{inputs}

\begin{figure} 
\begin{centering} 
\begin{tabular}{|l|c|rr|cc|} \hline  
{\bf Name}              &  {\bf Embedding} &  {\bf $n$}    & {\bf $m$ }        & {\bf $q$}      &  {\bf $L=q/n$}  \\  \hline  \hline 
CBFM              &  native       & 5387    &38751     & 5387    & 1.0 \\  
NAT1              &  native       & 5387    & 38751    & 5387   &  1.0\\
NAT7              &  native       & 5387    & 38751     & 5387   & 1.0 \\ 
TILE                &  native       & 5387    & 38751    &  5387  & 1.0 \\
FCL                 &   native       & $\overline{5029}$ & $\overline{22030}$  &   $ \overline{5029} $  &  1.0 \\  \hline 
3DLAT            & custom       &  2688&   7444       &  5376 &  2.0\\    \hline   
BPSP             & heuristic    & $\overline{867}$  & $\overline{1211}$ &   $\overline{3037} $  & $\overline{3.4}$ \\  
DREG03    &  heuristic    & 754           &  1131                                   &  $\overline{2856}$   &  $\overline{3.8 }$\\   
SOCs        & heuristic     & 355          &  1053                                    &   $\overline{1263}$  & $\overline{3.4}$ \\     
SOCu            &  heuristic    & 355  & 1053                                        &   $\overline{1263}$   & $\overline{3.4}$ \\  \hline 
SK                &   clique & 175    &  15225      &  2625 & 15   \\  
CDMA          &  clique & 175     & 15225         &  2625  & 15 \\ 
DAIG   &   clique  &  $\overline{174}$   &  $\overline{15051}$         &  2625  &  $15$  \\  \hline 
\end{tabular} \\
\end{centering} 
\caption{Our testbed of input classes.  Input sizes (nodes $n$ and edges $m$) correspond to the largest graph of 
each class that can be reasonably embedded onto the QPU.   Column $q$ shows the number of qubits used in physical (embedded) versions of these graphs.  In the last column, $L$ denotes mean chain length.  This is the mean ratio of physical to logical problem sizes: for example, physical SK inputs have 15 times more variables than logical  SK inputs.  Numbers with overlines are averages taken over the 15 randomly generated graphs or embeddings in a given class.} 
\label{fig::inputlist} 
\end{figure} 

Figure \ref{fig::inputlist} summarizes the 13 input classes that were selected for our study.  Columns labeled $n$ 
and $m$ show the number of nodes and edges in logical (pre-embedded)
graphs;  column $q$ shows the number of nodes in the physical (post-embedded) graphs.  The last column shows the mean ratio of nodes in embedded versus unembedded versions of each graph category, equivalent  to mean 
chain length.  Inputs are listed in decreasing order by problem size $n$, and are grouped by the strategy used for embedding, as follows. 
\begin{enumerate}
\item[\bf native] Five input classes are generated directly on the Pegasus hardware graph in the Advantage QPU;  their logical and physical versions are identical.  These are the largest inputs in our tests, using nearly all available 
qubits and couplers.\footnote{The Advantage\_system4.1 QPU in our tests contains a small proportion of unused qubits and couplers that did not meet quality standards and were disabled when the system came online.  
Active qubits and couplers in the hardware graph number 5627 and 40279, representing yields of 99.77\% and 99.49\%  respectively, of the full Pegasus graph.  Our tests use a subgraph of that hardware graph because 
of a design limitation of our GPU-based native solvers, which ignore some irregularly-connected qubits on the periphery.}  

These inputs use weighting schemes $(h,J)$ that have been identified as challenging for classical and quantum solvers to cope with.  The labels refer to {corrupted biased ferromagnets} (CBFM) from \cite{tasseff22}; {native spin glasses} (NAT1, NAT7) discussed in \cite{katz14}; {frustrated tiles} (TILE) \cite{perera20}; 
and {frustrated cluster loops} (FCL) from \cite{jking19}.  

\item[\bf custom]  Graphs with regular lattice-like structures can exploit
efficient custom embedding techniques.    
Three-dimensional lattices (3DLAT), consisting of unit cubes that share corner nodes, arise in scientific applications and simulations \cite{harris3d}. 

\item[\bf heuristic]  Four input classes with random and irregular connection structures are embedded using the heuristic 
embedding tool {\tt minor\_miner} available in D-Wave's Ocean SDK \cite{docminorminer}.  Labels refer to
inputs for the Binary Paint Shop Problem (BPSP), similar to those in \cite{bpsp}; random 3-regular graphs (DREG03); and signed and unsigned social network graphs (SOCs, SOCu) \cite{networkx}.   

\item[\bf clique] Three input classes are embedded using the {\tt busclique} clique-embedding tool available in Ocean SDK.  They are: Sherrington-Kirkpatric (SK) graphs studied in statistical physics 
\cite{venturellisk};  inputs for the Code Division Multiple Access (CDMA) signal decoding problem;  and 
dense AI graphs (DAIG), generated by a tool that``learns'' hard substructures of multiple application problems.
This tool constructs inputs  such that $n$ is always even.  
\end{enumerate} 

\paragraph{Screening Inputs for Hardness} 
\label{seeking} 
Given an input graph $G$ with $n$ nodes and $m$ edges, the  {\em problem size} $n$ is the number of  
variables that are assigned solution values, and {\em input size} $k=n+m$ is the number of graph components (node and edge weights) required to fully specify the problem via formula (2).  Problem size is a key indicator of input hardness, in the sense that the space of solutions to be explored grows as $2^n$.  Input size gives an upper bound on embeddability,  since it must hold that $n + m \leq q +c$. 

We performed a pilot study to identify interesting input classes for our benchmark tests, prioritizing hardness, structural diversity, and application relevance.  Our tests require inputs that are both 
small enough to embed onto the Pegasus hardware, and large enough that 
Milestones 1 and 2 are at least feasible to demonstrate.  That is, if the problem is so small or easy 
that classical heuristics can regularly find ground states in less than the 16 ms lower bound on programming time --- before the quantum computation has a chance to begin --- then there is no point in performing the test.  We argue that too-small or too-easy inputs are not interesting candidates for quantum solution methods because they can be quickly solved classically, leaving little room for improvement.   

The native, custom-embedded, and clique inputs passed our screening criteria. We included three clique-embedded input classes to represent this important boundary case; among embedded graphs, these are the smallest in terms of problem size $n$ but the largest in terms of input size $k$. 

For the remaining heuristic-embedded inputs, we applied a two-step screening procedure as follows.    
\begin{enumerate}
\item  We used the heuristic embedder to find a largest {\em reasonable} input size, 
selecting $K$ such that the median value (over 3 random trials) of maximum chain length (per trial) is in the range $15 \leq L_{max} \leq 20$, slightly higher than found in clique embeddings.
 
\item The Greedy and SA heuristics were run on these largest input;  if they could find putative ground states --- as indicated by agreement on the same minimum energy over many random trials --- within the 16 ms lower bound on programming time,  the candidate were omitted from our main tests.  
\end{enumerate}  
As it turned out, the inputs rejected by this screening procedure were of sizes $175 < n < 355$,  both smaller
and denser than those that passed.\footnote{These included four variations on random Satisfiability inputs, and random k-regular graphs with $k \geq 10$.}  Benchmark tests on moderately dense graphs must await availability of larger and more-connected hardware graphs. 

We consider the fact that eight embedded graph classes {\em were} identified as test candidates to be an encouraging sign of progress, since (with a few exceptions mentioned in Section \ref{previous}) 
previous-generation QPUs have been too small for logical inputs of embeddable size to be considered viable candiates for the benchmarking tests in this paper.  
\subsection{Solvers}
\label{solvers} 
 
The table in Figure \ref{solverslist} lists the heuristic solvers in our study.    
The first three read only physical inputs defined on the Pegasus hardware graph, which may be either 
native or post-embedded versions of logical graphs.  The remaining five solvers read general graphs, 
which can be either pre-embedded logical inputs or native inputs.  Both physical solvers and one general solver (PTg) are implemented on GPUs;  the remaining four general solvers are implemented on CPUs.  

\begin{figure}
\begin{centering}
\begin{tabular}{|llcc|}\hline 
Name                                        & Label              &   Inputs     &  Platform    \\  \hline \hline
D-Wave {\tt advantage\_4.1}                  & QPU  &   physical   &  QPU \\   \hline 
Simulated Annealing-Optimize  & SAo\_native   & physical   &  GPU \\ 
Simulated Annealing-Sample     & SAs\_native   & physical   &  GPU \\   \hline  
Random                                     & Random        &  general   &  CPU  \\ 
Steepest Greedy Descent            & SGD              &  general    &  CPU  \\   
Simulated Annealing (Logical)     & SA                &  general   &  CPU  \\
Parallel Tempering (sequential)   & PTc             & general    &  CPU  \\  
Parallel Tempering (parallel)       & PTg              &  general    &  GPU  \\   \hline 
\end{tabular} \\
\end{centering} 
\caption{Solvers in our benchmark study.  The third column indicates whether the solver reads 
 physical or general inputs.  The fourth column shows the type of platform on which the solver runs.}
\label{solverslist}
\end{figure} 

\paragraph{Classical solvers} 
The classical solvers all implement variations on the {heuristic search} paradigm;  these are 
sometimes called landscape-traversal methods, based on a popular metaphor that envisions the solution space as a {landscape} defined over all possible solutions $x$, where the surface elevation at point $x$ corresponds to the objective function value defined by formula (2).  A landscape-traversal solver starts at a random solution point $x$ and then moves step-by-step through the landscape, making incremental changes to $x$ by flipping one (or more) binary variables, while seeking lower ground.  A {\em local minimum} is a valley surrounded by uphill moves; a {\em global minimum} is 
a valley of minimum elevation. 

{Total work} $w$ corresponds to the number of iterations performed while moving 
about the space;  this is an upper bound on the total number of states visited, since some solvers can repeat visits to the same states.  More work is associated with longer computation time (the exact
relationship depends on the implementation), but tends to produce better-quality solutions.  Different solvers manage this tradeoff differently, as follows. 
\begin{itemize} 
\item The {\bf Random} solver generates an initial solution and makes no effort to improve it, so $w=0$.  This 
solver provides baseline measurements of the fastest runtimes and worst-quality solutions among landscape-traversal heuristics.  

\item A greedy solver moves strictly downhill.  At each iteration, the steepest gradient descent ({\bf SGD}) solver studied here considers all single-variable flips and selects the one giving the largest improvement in (2), stopping and reporting the result when no improvements can be found.  Work $w$ is determined by the  number of visited states between a random start and a nearby local minima, and solution quality reflects the distribution of local minima reachable from random starts. 

\item Our simulated annealing ({\bf SA}) solver has total work controlled by the user via a {\tt num\_sweeps}  parameter.  The {sweep} loop iterates over $n$ nodes when deciding where next to step, so that $w =n\cdot{\tt num\_sweeps}$.  A temperature parameter $\tau$ is used to probabilistically decide which node is used for the next-step, which may be uphill; $\tau$ decreases with sweep count so that uphill moves become less likely over time.  This approach allows SA to escape local minima earlier in the computation, becoming more strictly greedy in later stages. 

\item The {\bf SA-native} solver is a GPU-based version of SA that parallelizes the sweep loop by 
assigning $n$ nodes to $n$ CUDA cores, so that the computation of next-step probabilities can exploit $n$-way parallelism. Note that the step itself is not parallelized because the GPU can only be in one state at a 
time and computation of next-state probabilities depends on the previous state;  that is, parallelization speeds up time-per-sweep, but the work/iteration loop is inherently sequential since the 
computation can only be in one state at a time.   
This solver reads only Pegasus-structured inputs.  We implemented two versions: for {\bf SAo\_native}, parameters are set to prioritize optimization (more rapid descent to a low-energy solution), and for {\bf SAs\_native}, parameters are set to prioritize sample diversity (less likely to be stuck in local minima).   

\item Parallel tempering ({\bf PT}) is based on SA but contains an extra loop to represent $R$ {\em replicas} of SA, each working on its own solution $x_r$ and operating at a constant temperature $\tau_r$.  At intervals, solutions are swapped among adjacent replicas (i.e.,  the replica at $\tau_r$ swaps with $\tau_{r+1}$ and $\tau_{r-1}$) to move the best candidate solutions toward lower $\tau$.  
The user parameter {\tt num\_sweeps} controls the total number of sweeps for all replicas.  Over time, the number of 
replicas can grow or shrink according to swapping frequencies; for a mean number of replicas $\overline{R}$, total 
work is equal to $w = \overline{R}\cdot n \cdot {\tt num\_sweeps}$.  We tested two implementations of PT:  {\bf PTc} runs sequentially on a CPU;  {\bf PTg} parallelizes computations in the sweep and the replica loops, and runs on a GPU.   
\end{itemize} 

\paragraph{Quantum annealing solver} 
See \cite{ieee, cross-dis} for introductions to computation by quantum annealing.  Briefly, the QPU control system  
follows a time-varying Hamiltonian ${\cal H}(t)$ over a time interval $t: 0 \rightarrow t_{anneal}$,  to create a smooth transition from an {initial} transverse field Hamiltonian ${\cal H}_{initial}$  to a {problem} Hamiltonian ${\cal H}_{problem}$ that corresponds to the objective function (2).  The qubits behave like a quantum particle process driven 
by ${\cal H}(t)$, naturally seeking their collective ground state, which corresponds to an optimal solution to (2) at the end of the anneal. 

In landscape-traversal  terms, the computation can be visualized as taking place on an initially flat landscape (the transverse field), from which the problem landscape defined by (2) gradually emerges.  Rather than stepping point-by-point through this moving landscape, the qubits exploit quantum superposition and entanglement to probabilistically represent ``all states at once,''  in such a way that highest probabilities track lowest-lying areas of the landscape as it evolves.  At the end of the anneal, the classical problem Hamiltonian ${\cal H}_{problem}$  dominates ${\cal H}(p)$:  classical state is read according to those final probabilities and returned as the solution. 

Note that our classical conception of work does not translate well to the quantum computation because qubits do not visit 
states one-by-one; indeed there are no iterations to be counted.  Instead, we modify our definition of work to match 
the natural quantum unit of computational effort, and set $w = t_{anneal}$.  

In an ideal noise-free environment, if  $t_{anneal}$ is above a certain threshold time 
determined by the input, the qubits will finish in a ground state of the objective function (2) with high probability \cite{farhi-science, nishimori-qa, albashtheory}.  In a real-world quantum system, noise and imprecision of the analog control system can cause the qubits to move away from their ground state and instead finish in a low-energy state of (2) \cite{albashtemp, young}.  

\subsubsection{\bf Fair Test Policy} 
\label{params} 

The QPU, SA, and PT solvers offer several runtime parameters to the user, which can be tuned to elicit best performance on specific input types.  The question of how to design a so-called {\em fair test}, which ensures equal parameter-tuning effort across all solvers,  has received much attention in the optimization methodology literature \cite{barres, dunning, hooker, dsj, mcgbook}.  The goal of fair testing is to ensure that outcomes can be replicated by practitioners;  Johnson \cite{dsj} recommends that all tuning procedures be well-defined and algorithmic, with tuning time included in reported runtimes.  

Following this advice, we implemented a wrapper code for each solver that accepts exactly three inputs --- the 
instance ($G$ or $P$),  the sample size $s$, and the time limit $t$ --- and sets solver parameters to either fixed defaults or auto-tuned values according to code-based policies, with auto-tuning included in measured runtimes.  The most important parameters, controlling work,  were auto-tuned as follows.  

\begin{itemize}
\item For Random and SGD, which offer no user parameters, the wrapper code simply takes a maximum
number of samples within the time limit $t$.  A  sample of $s$ lowest-energy results is subsampled from the full set 
during data analysis.  

\item  For the SA and PT solvers, best results come from maximizing work to meet $(s,t)$.  The wrapper code sets the {\tt num\_sweeps} parameter to the maximum possible within the per-solution time limit $t/s$, according to quick time-per-sweep estimate that is performed once per solver per instance and re-used for all combinations of $(s,t)$.  
A separate procedure occasionally checks progress against the full time limit $t$ and makes adjustments to {\tt num\_sweeps} if necessary.  

\item For the QPU, maximizing anneal times to meet test limits of $t \in [20 \ldots 200]$ ms is not possible under 
current (public access) policies, which support anneal times in range $t_{anneal} \in [.0005, \ldots, 2]$ ms.  Furthermore, 
we do not  necessarily expect best results from maximizing this parameter:  experience suggests a law of 
diminishing returns whereby setting $t_{anneal}$ to higher values within its range yields no perceptible improvements in solution quality. 

Therefore our wrapper code aims to maximize {quantum utilization} with a default anneal time set 
to $t_{anneal} = t_{readout}$, and default $r$ set such that $t_{program} = r (t_{anneal} + t_{readout})$. 
This time block is multiplied to meet a given test scenario $(s,t)$, with reprogramming steps as described in Section \ref{accessov}.  If this default setting does not allow enough time for the required number of samples, then reads 
are increased and programming steps are reduced; if this is still not, anneal time is reduced.  As with Random 
and SGD, if the QPU returns more samples than needed in a given scenario,  the $s$-best solutions are subsampled during analysis. 
\end{itemize} 

\subsection{Performance measurement and analysis} 

We use one common test procedure to gather data for both milestones as follows.  
For each of 13 input classes we generate 25 random instances.  
A native input has one version that is sent to both logical and physical solvers; a logical input has a pre-embedding version sent to logical solvers  (Random, SGD, SA, PTc, PTg) and a post-embedding version sent to physical solvers (QPU, SAo-native, SAs-native). 

For each input instance and each solver,  we construct a test scenario $(s,t)$, requesting $s = [1,10,100,1000]$ solutions to be returned within time limit $t=[.02, .05, .1, .2, .5, 1]$ seconds.  Of 24 possible scenarios, we omit five because $t/s$ is below the (amortized) lower bound on QPU access time;  thus, $t/s$ is in range $[.0002 \ldots 1]$ seconds in our tests.  

In Milestone 1 tests using 13 input classes, all solvers read identical inputs: we compare QPU performance to logical solvers on the five native classes, and to physical solvers on all classes, and solution energies are 
calculated in the physical problem space.  In Milestone 2 tests using eight logical inputs, the five logical solvers
read pre-embedding inputs while the QPU reads their post-embedding physical versions.   The physical solutions 
returned by the QPU are mapped back to the logical solution space for evaluation;  broken chains are resolved using the default parameter setting {\tt chain\_voting = True}.  

Statistics of sample solution quality are calculated as follows.    
\begin{itemize}
\item Let ${\cal S} (a, x, s, t)$ be a sample of $s$ energies returned by solver $a$ for instance $x$ when run for time limit $t$.  Sample solution quality is measured by {\bf median sample energy}, denoted ${\cal M}(a,x,s,t)$.  

\item For each instance $x$, the {\bf target energy} ${\cal T}(x)$ is the minimum energy observed 
in all tests over all solvers and scenarios.  This is not necessarily the optimal energy for $x$.  

\item The {\bf relative error} for solver $a$ on instance $x$ is the absolute scaled difference between 
${\cal M}(a,x,s,t)$ and ${\cal T}(x)$, 
\begin{eqnarray}
{{\cal R}}(a,x,s,t) & = &  {\left| {  {\cal T}(x)  -  {\cal M}(a,x,s,t) }  \right| \over \left| {\cal T}(x)  \right|}. 
\end{eqnarray}  
\end{itemize} 
Relative error is equal to zero when at least half of the sample energies found by $a$ on input $x$ 
are equivalent to the target energy.  When relative error is nonzero,  ${\cal R}(a,x,s,t)$ indicates 
how close the solver came to finding target energies. 

\section{Results} 
\label{results} 

\begin{figure}
\begin{centering}
\includegraphics[width=6in]{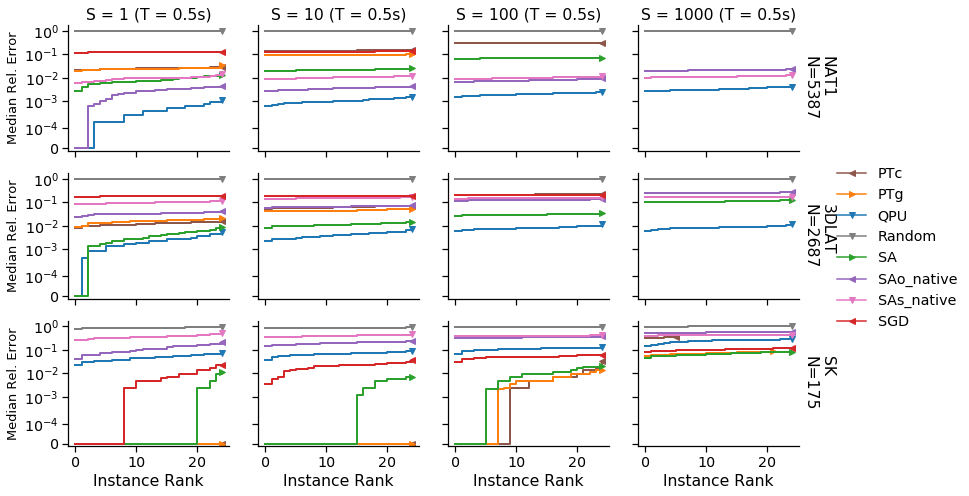}
\end{centering} 
\caption{Performance results for the NAT1 (top), 3DLAT (middle) and SK (bottom) input sets,  
for time limit $t=.5$ seconds and sample sizes $s = [1,10,100,100]$.  
Each panel shows a plot of median relative error for 25 inputs from the set, sorted in increasing order. 
These are symlog \cite{symlog}  plots with $y$ on a logarithmic scale that 
switches to linear to expose the critical region near $y=0$:  the apparent step functions are due to the switch of scale and not indicative of discontinuities in the data.   } 
\label{fig::readcactus} 
\end{figure}

Figure \ref{fig::readcactus} shows example results for three input sets from our tests:  NAT1, 3DLAT, and SK.   These inputs  all have spin-glass weight with fields $h_i =0$ and couplings chosen u.a.r. 
from $J_{ij} \in \{ -1, +1 \}$.  They represent three extremes of graph structures and sizes:  the native Pegasus graph (NAT1, $n=5387$), among the largest tested;  three-dimensional lattices (3DLAT, $n=2688$), the largest embedded graphs;  and Sherrington-Kirkpatric graphs build in cliques (SK, $n=175$), among the smallest of embedded graphs.   

Each row shows results for four test scenarios,  corresponding to sample sizes $s=[1,10,100,100]$ at the 
time limit $t=.5$ seconds.  Each panel shows curves for median relative 
error ${{\cal R}}(a,x,s,t)$ over 25 instances, sorted in increasing order and  color-coded by solver:  the rightmost point is the worst $\cal R$  returned by the solver over all 25 instances, and the midpoint $x=13$ is the median result over all instances.  

These are {\em symlog} plots \cite{symlog}  with a logarithmic y-scale that becomes linear near $y=0$, which 
give the best view of outcomes in our rank-based analysis;  the steps at low $y$ are due to the scale change and 
are not indicative of significant discontinuities in the data.   

For each input class and scenario, we evaluate solver performance as follows.  
\begin{itemize} 
\item A {\em win} is awarded to a solver if its ECD curve is strictly below those of all other solvers on at least half (13) of the instances.  For example, the QPU wins against all solvers on all four NAT1 scenarios.  These wins are recorded for Milestone 1 because all solvers read the same physical inputs.  

A {win} is also awarded to a solver that ties with other winning solvers on at least half the instances.  For example, in the first SK
scenario ($s=1$),  SA, PTc, and PTg share the win, but SGD does not because it ties in fewer than 13 instances.  

A win for Milestone 1 is awarded to a physical solver that outperforms other physical solvers reading embedded problems.  For example,  the QPU outperforms both SAs\_native and SAo\_native in all SK scenarios. 

\item A {\em fail} is awarded if a solver cannot return all requested samples within the time limit, on at least half the inputs tested; if it fails on all inputs the ECD line is absent.  For example, in the fourth NAT1 panel with $s=1000$, all  logical solvers fail on all 25, and the three physical solvers succeed.  
\end{itemize}
A solver that neither wins nor fails in a given scenario is said to compete. 


\begin{figure}
\begin{centering}
\includegraphics[width=.49\textwidth]{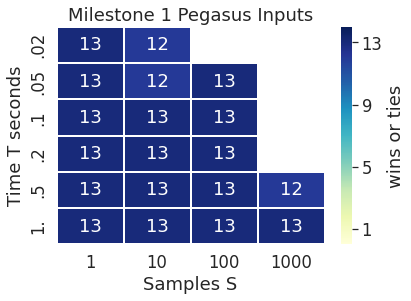} 
\includegraphics[width=.49\textwidth]{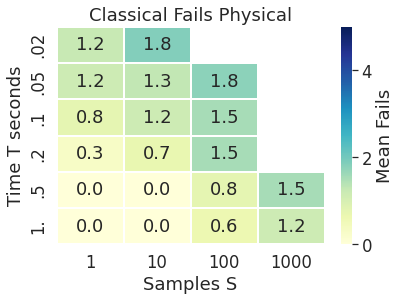} \\
\end{centering}
\caption{Panel (a):  Number of wins or ties earned by QA in Milestone 1 tests, 
over all scenarios and (physical) inputs.  The 
maximum number of wins per scenario is 13,  the number of input classes tested.  
Panel (b): number of classical failures in tests using physical inputs, averaged over all input classes.  The maximum number of failures per input class is seven or two, equal to the total number of solvers tested.} 
\label{fig::m1} 
\end{figure}

\subsection{Milestone 1:  Physical Solvers and Inputs}
\label{resultsm1} 

Figure \ref{fig::m1} summarizes Milestone 1 results, which consider cases where all solvers read the same 
physical inputs:  QPU versus logical and physical solvers on eight native inputs, and versus physical solvers on five embedded  inputs.   
Panel (a) shows tallies of QPU wins over all 13 input classes in 19 scenarios, comprising 475 tests.  The QPU clearly dominates classical 
approaches, winning in all but 3 of 475 tests covering 13 input classes and 19 scenarios (98.5 percent).   

Panel (b) shows a tally of number of classical fails per input class in each scenario, 
considering the five native inputs on which all 
seven solvers were tested.   The number of possible failures in any test scenario is between 0 and 7;  each box 
shows the mean number of fails per scenario, averaged over all five native inputs.  

Here are some details. 
\begin{itemize}  

\item Nearly all of the QPU wins are solo wins: a few ties with SAo\_native and SAs\_native were found in scenarios with large $t/s$ (lower left corner), on the smallest (clique) inputs in our tests.    
 
\item The three cases where the QA does not win are near-ties versus SAo\_native on CDMA inputs.  These 
inputs are distinguished from others in our testbed by containing unusually high-precision Gaussian weights $J,h$.
Higher precision poses a challenge to quantum annealing due to finer gradations of elevation in the solution landscape, which are
harder for the QPU to distinguish.    

\item  Considering only classical solvers, the  GPU-based SAo\_native and SAs\_native solvers performed best overall,
confirming the intuition that parallelism can boost performance of the SA 
heuristic.\footnote{GPU solvers implemented in C++ and CUDA, and run on an NVIDIA GeForce GTX 1080-Ti 
processor with 3584 CUDA cores and 1582MHz processor clock speed.  All CPU tests ran on a 
3.30GHz  Intel Core i9-7000x CPU with 20 cores.}   It is interesting to note that on these 5000-node 
Pegasus inputs, GPU parallelism yields only about a 60-fold speedup over comparable CPU solvers;  on 
previous-generation Chimera inputs, GPU solvers saw 133-fold speedups.  This degradation in speedup relative to sequential solvers can be attributed to the fact that parallel speedup is reduced by increased graph connectivity: the Pegasus graph is nonbipartite with 
degree 15, whereas Chimera is bipartite with node degree 6.  

\item The upper right boundary where classical solvers are most likely to fail corresponds to small $t/s$, 
below about 0.5 milliseconds.  The physical solvers (QPU and SA\_native) never failed in any scenario. 
Among the logical solvers, PTc and PTs failed most frequently.   
 \end{itemize}  

\subsection{Milestone 2: Logical Solvers and Inputs} 
\label{resultsm2} 

\begin{figure}
\begin{centering}
\includegraphics[width=.4\textwidth]{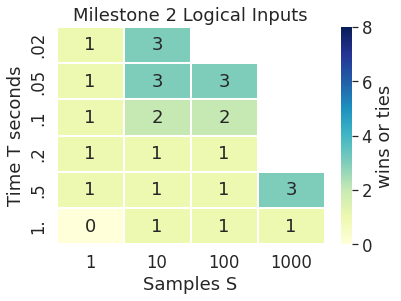} 
\includegraphics[width=.4\textwidth]{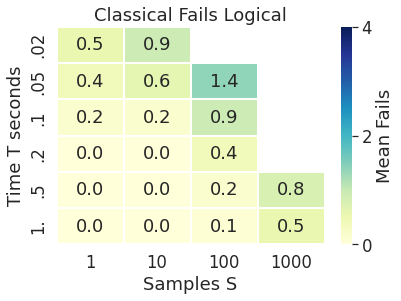}\\
\end{centering}
\caption{Milestone 2.  Panel (a) Number of QA wins (out of eight maximum) over all logical inputs and scenarios. 
Panel (b) is the mean number of classical fails (out of five maximum), averaged over eight input classes.} 
\label{fig::m2} 
\end{figure}

In Milestone 2 tests we consider performance on eight general input classes, for which five logical solvers read smaller pre-embedding versions and the QPU reads larger post-embedding versions of the same problems.  
(Performance of physical solvers versus the QPU are considered in M1 test).   Thus, at most eight wins and at 
most five fails can be awarded per scenario.  

Results appear in Figure \ref{fig::m2}.  Panel (a) shows that the QPU wins  in 28 of 152 cases, or about 18\% of problem scenarios.  In other cases, the QPU competes (neither wins nor fails).  Panel (b) shows the mean number of classical fails per scenario per input class.  Although classical solvers do a better job overall in this milestone,  it is interesting to note that the scenarios where classical solvers can fail are all on the upper right boundary, just as in Figure \ref{fig::m1}.   Panel (a) shows that these are the same scenarios where the QPU solver is most likely to win.  
Here are some details.  

\begin{itemize}
\item  The QPU is more likely to win, and classical solvers more likely to fail, 
in tests involving small $t/s$.  Furthermore, 
these outcomes are strongly correlated with problem size.  In Milestone 1, QPU wins nearly always 
when $n \geq  5029$.  In Milestone 2,  QPU wins occur on the three 
largest embedded input graphs:  3DLAT ($n=2688$)  in 18 scenarios, BPSPh ($n=867$)  in 6 scenarios,  and  DREG03 ($n=754$) in 4 scenarios. 

\item Among classical solvers, SA was the sole winner in most cases.  The exception is scenarios with large $t/s$  (lower left corner) and small $n$ (e.g. clique problems),  for which many solvers shared the win.   PTc and PTg  failed most frequently at small $t/s$, as in M1 tests.  

\item  The GPU-based PTg solver never outperformed its CPU-based counterpart in these tests; sometimes they 
tied, and sometimes PTc outperformed PTg.  We attribute this outcome to the fact that algorithms for irregularly structured graphs are notoriously challenging to parallelize effectively \cite{biswas}. 
\end{itemize} 

\section{Discussion: What drives performance?}
\label{discussion}

\begin{figure}
\begin{centering}
\begin{tabular}{|l|rrcc|} \hline 
           & Problem &Edges  & Input Size      & Chain Length  \\ 
Graph   & Size $n$ & $m$  & $K=n+m$                     & $L=q/n$  \\ \hline \hline
Native   & 5387  & 38751  &   44138                         &  1.0  \\ 
3DLAT  & 2688  & 7444    &   10132         &  2.0\\ 
Clique    & 175   &  15225 & 15400         & 15 \\ \hline
\end{tabular}
\end{centering}
\caption{Sizes of three graph structures used in our tests.    
}
\label{fig::embeddings}
\end{figure} 

This section considers relationships between input properties and performance of classical and quantum 
solvers that explain some of the observations of Section \ref{results}.  For simplicity we focus on three graph structures (NAT1, 3DLAT, and SK),  and on the Advantage QPU versus three classical solvers (Random, SGD, and SA), all implemented on CPUs.  (The PTc solver, not shown here, was dominated by SA on NAT1 and 3DLAT inputs and by SGD on SK inputs.)

Figure \ref{fig::embeddings} shows some relevant size metrics for these graphs. Given a graph with $n$ nodes and $m$ edges,
\begin{itemize}
\item {\bf Problem size} $n$ is the number of variables,  which 
determines the size $2^n$ of the {solution space} to be searched, a key indicator of problem complexity. 

\item {\bf Input size} $K=n+m$  is the number of components (node and edge weights) needed to fully 
specify the problem.  This correlates with the size of a key data structure (and a lower bound on computation time) for 
classical solvers, and with the size of the embedded input on the QPU.  

\item  {\bf Chain length} $L=q/n$ is the mean ratio of problem size in logical and physical versions of each 
instance (although technically speaking native graphs do not contain chains).  Chain length serves 
to quantify the indirect overhead cost of embedding that is considered in our Milestone 2 tests. 
\end{itemize} 
In terms of problem size, NAT1 is among the largest graphs in our study, 3DLAT is the largest embedded graph, 
and SK is among the smallest of embedded graphs.  3DLAT is about 15 times larger than SK in problem size, 
while SK is about 50\% larger in input size, with 7.5 times longer chains.  

\begin{figure}
\begin{centering}

\includegraphics[width=.8\textwidth]{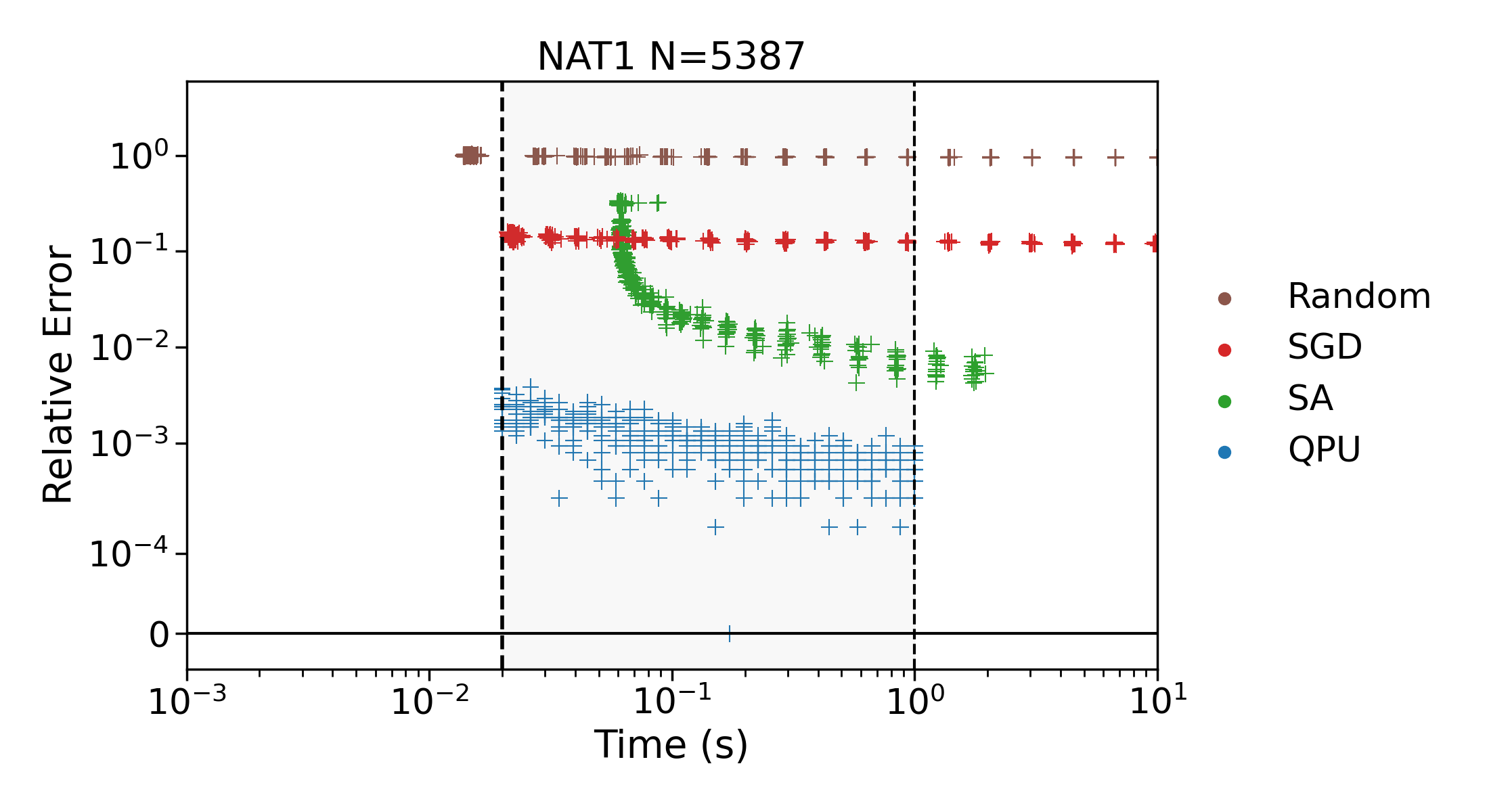} \\
\includegraphics[width=.8\textwidth]{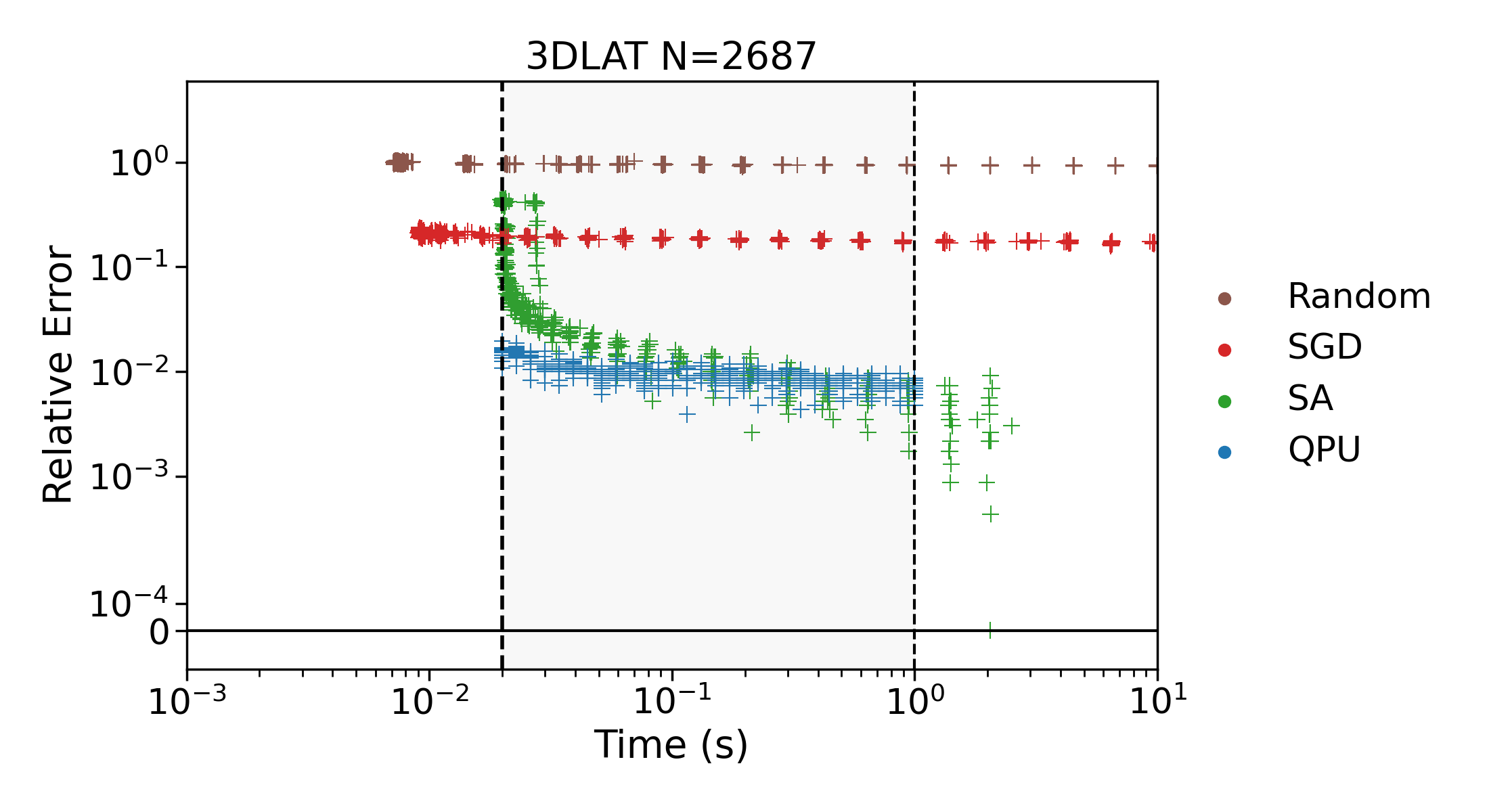} \\
\includegraphics[width=.8\textwidth]{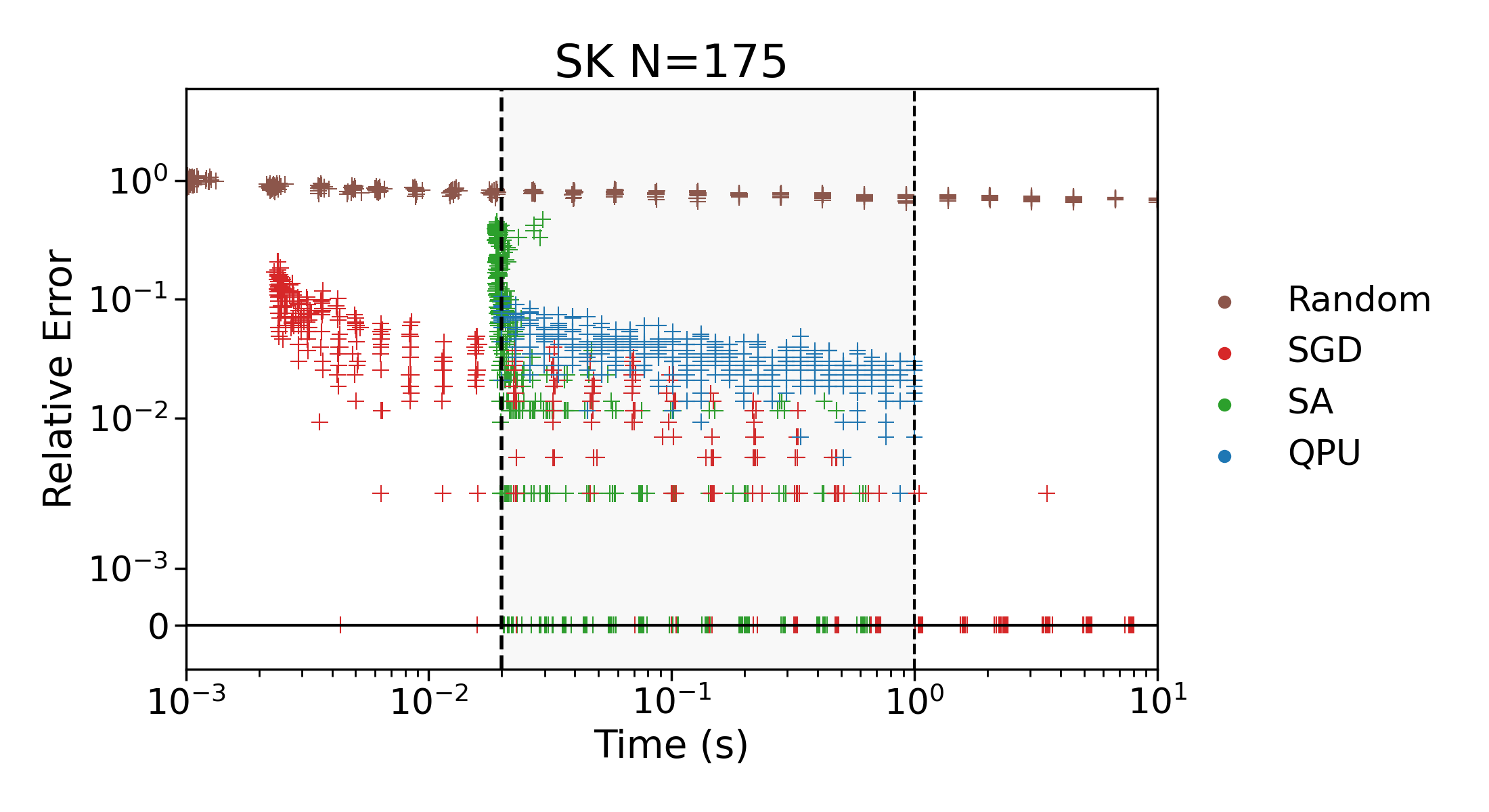} \\

\end{centering} 
\caption{Convergence of minimum relative error with respect to computation time, for three 
inputs and four solvers.  These panels show results for 15 independent trials for time limits $t$ in geometric increments.  The shaded regions mark the range of test times from Section \ref{results}.} 
\label{fig::converge}
\end{figure}

Figure \ref{fig::converge} shows how solution quality converges towards optimality with 
increasing computation time,  with work controlled by the auto-tuning policies described in Section \ref{solvers}.  
The three panels are ordered bottom-up by increasing problem size.
In each panel, the y-axis marks relative error ${\cal R}_m$ observed in 15 independent trials 
at geometrically-increasing time increments $t$.  Note that cases where the solver collects multiple samples to fill the time limit, relative error here is the {\em minimum} energy found in the batch, not the {\em median energy} as measured in Section \ref{results}.  The shaded area marks the range of time limits considered in that section.

These results showing sample minimums are consistent with results for sample medians in 
Figure \ref{fig::readcactus}.  On NAT1 (Milestone 1),  the QPU outperforms classical solvers in all 
scenarios,  and SA fails in scenarios where time per sample is low (here when  $t/s < .006$ seconds).    
On 3DLAT (Milestone 2), the QPU outperforms classical solvers when $t/s < .1$ sec;  
SA can win when $t/s > .2$ sec, but can fail when $t/s < .02$ sec.  On SK inputs (Milestone 2), QPU results are 
dominated by multiple classical solvers;  but SA can fail when $t/s < .02$ sec.  

These three inputs have {spin glass} weights, with fields $h_i =0$ and edge weights 
selected u.a.r. from $J_{ij} \in \{ -1, +1 \}$;  all solution energies are separated by multiples
of $\pm 2$.  Results for 3DLAT and SK are shown in the logical problem space, after physical solutions from the QPU 
have been mapped back to the original input. 
 
\subsection{Classical Performance Drivers}
Recall from Section \ref{solvers} that the classical solvers in our tests work to improve an initial random solution by flipping one node at a time, while seeking lower-energy regions of the solution space. The work parameter $w$ equals the number of main-loop iterations performed as a solver explores the space, which is an upper bound on the total number of states visited.  We consider the relationship between work, time, and solution quality for these solvers.  

First, Figure \ref{fig::converge} illustrates the normal expectation that computation time per unit of 
work increases with input size $K$.  Moving from bottom to top, the classical start-times (leftmost data points) 
generally shift right, reflecting increasing costs to initialize data structures, autotune parameters (if applicable), 
and return the first solution.  For SGD and SA, both start-time and time-per-iteration are proportional to $K$ 
because these operations require full traversals of the input data structure.  

Second, the three classical solvers form a Pareto frontier ranging from least-time/worst-solutions to most-time/best-solutions as follows.
\begin{itemize}
\item  Random applies no work to improve the initial solution ($w=0$).  Here, ${\cal R}_m$ follows the 
extremal statistics (minima) of random solution samples that increase in size with $t$.  
(Note that the distribution of solution energies for spin-glass inputs is symmetric around zero:  therefore, minima of small random samples are near zero, which by formula (3) works out to ${\cal R}_m \approx 1$ in all three panels.)    

\item  SGD does more work than Random, resulting in later start times and better-quality solutions.  For any 
specific input, mean work per solution is constant, reflecting the average traversal distance from a random initial 
solution to  a nearby local minimum.  As with Random, the data points follow the extremal statistics of increasingly large samples from the distribution.  

\item The SA work parameter {\tt num\_sweeps} is set to maximize work-per-solution while respecting time 
limit $t$;  increasing work allows SA to climb out of local minima and continue its downhill progress.  
\end{itemize}

Compared to the barely perceptible improvement of Random and SGD in the top two panels, the SA strategy 
of increasing work is clearly a more effective use of the alotted time. The difference is that increasing work allows 
SA to drive the center of {\em location} of its sample distribution toward lower energies, whereas Random and SGD are stuck sampling from fixed distributions created by constant work.  

Also in  the top two panels, we note that the gradual flattening out of the SA convergence curve --- in the sense that doubling work is much less effective at large $w$ than at small $w$ --- is not unexpected in the context of NP-hard 
problems.  The existence of an approximation method that achieves polynomial improvements in solution 
quality with only polynomial extra work in the worst case (called an FPTAS), would imply that $P=NP$ \cite{vanLeeuwenfptas}. 


The bottom SK panel runs contrary to these expectations, since both SGD and SA converge 
quickly to putative optimal solutions, as indicated by agreement on the same minimum 
energy (at ${\cal R}_m = 0$) over many independent trials.  In this case, SGD has no problem finding ground 
states among the local minima that it samples, and SA needs only a few sweeps to routinely find ground states.
(Figure \ref{fig::readcactus} shows that PTc and PTg also performed well on SK inputs at higher $t/s$.)    

This result is not surprising in the sense that any NP-hard problem class can contain easy inputs that are 
amenable to greedy and heuristic search approaches.  Although SK inputs are hard enough to pass our screening tests of Section \ref{inputs} (needing more than 16 ms to find ground states in their sample medians), strong performance from several classical solvers suggests that the problem landscape is smooth, with relatively few obstacles to impede progress toward ground states, or that ground states are unusually common, or both.  Determining whether 
these ``easy'' conditions are common to clique-like graph structures or are simply artifacts of small problem size, is an interesting topic for future study. 

\subsection{Quantum Performance Drivers} 
We now consider the relationship between work  $w = t_a$ and solution quality ${\cal R}_m$ on the 
Advantage QPU, and how that relationship is affected by input properties.    

Note that nothing definitive can be concluded based on the {\em shapes} of the three convergence 
curves in Figure \ref{fig::converge}, for two reasons.  First, by our auto-tuning 
policy (Section \ref{solvers}),  anneal time is $t_a = 240\; \mu$s throughout most of the test range:  these convergence patterns are due to increasing sample size, not increasing work.  
Second, the data points are plotted according to access times rather than anneal times, which distorts the true energy-versus-work relationship: an anneal-time plot would show QPU start times at $t_a = 240\; \mu$s, about 750 times lower than minimum access time (16 ms), and beyond the left edge of the plot field, thus demonstrating Milestone 0 on all three inputs.  

On the other hand, the {\em vertical locations} of QPU data points relative to classical results are informative:  QPU solution energies are dominated by both SGD and SA in the bottom panel,  intersect with SA in the middle panel, and lie 
strictly below the Pareto frontier in the top panel.  This reverse trend, by which QPU solution energies {\em improve} 
relative to classical energies on increasingly large inputs and under fixed time limits, outright contradicts any assumption 
that quantum computation time must necessarily increase with problem size to ensure competitive results.  

This divergence illustrates a fundamental distinction between quantum and classical models of 
computation.  As a popular quantum computing metaphor explains it: a classical solver can represent 
one of $2^{n}$ solution states with a register $R$ of $n$ bits that are updated incrementally as the solver moves 
from state to state.  In contrast, the quantum solver can operate on a register $Q$ of $n$ qubits, exploiting superposition and entanglement to probabilistically represent all $2^{n}$ states simultaneously, and manipulating $Q$ so that the most desirable states are most likely to be observed at the end of the computation.  Since the quantum computation does not visit states one-by-one, the quality of solutions it finds does not depend on how many states might be visited in a given 
time limit.   

Indeed, since the QPU always operates on the full Pegasus graph 
of size $q+c$, irrespective of how many qubits and couplers are used to represent an input of size $n+m$, there is no direct structural mechanism (such as a data structure being traversed) that would require quantum computation time to vary with logical problem size.  
   
Instead, models from statistical physics tell us that outputs from quantum annealing computations 
can be approximately characterized by a Boltzmann distribution with effective temperature $\tau$;  
lower effective temperature is associated with distributions returning better-quality solutions.  
Effective temperature depends on three factors:
\begin{itemize}
\item Physical properties of the QPU (such as environmental noise and fidelity of analog controls),  
which do not vary with input \cite{albashtemp, benedetti, nishimori, raymond, young};   

\item Annealing parameters (such as anneal time and location of pauses in the anneal path), which 
can have significant affect on solution quality \cite{goldenreverse,marshallPause, venturellireverse}, but (with two
exceptions discussed below) were not varied in our tests.  

\item Properties of the physical inputs that determine landscape structure, such as the 
range of weights $(h,J)$ \cite{katz14, king15, jking19, perera20, tasseff22}. 
\end{itemize}  

Recall that by our auto-tuning policy,  all annealing and input parameters were held fixed, 
with two exceptions: during each (re)programming step, a random spin-reversal transform (an annealing parameter) was applied to native inputs; and a random increment/decrement of chain strength (an input property) was applied to embedded inputs.  Each such randomization can shift the location $\tau$ of the Boltzmann distribution incrementally 
higher or lower:  thus, randomizing over multiple programmings has the effect of spreading the aggregate distribution 
over a wider range.  This explains the gradual downward trend in minimum relative error ${\cal R}_m$ with larger samples at increasing time limits, even though work is constant; however it does not explain the divergence in performance relative to classical solvers.  

Strong performance in our Milestone 1 tests suggests that whether a 
physical input is native or embedded, $\tau$ is consistently low enough for the QPU to 
outperform classical solvers on all inputs tested.   

We conclude that the observed pattern of quantum outcomes in Figure \ref{fig::converge} 
is due to a mechanism exposed by the unembedding step,  when physical solutions are mapped back to logical 
solutions and energies are calculated in the original problem space.  Specifically, we conjecture that quantum 
solution quality is negatively correlated with chain length $L = q/n$: chain length grows inversely with $n$ in our test design because embeddings range from large-and-sparse (with short chains) to small-and-dense (with long chains).   We have observed, for example, that longer chains require stronger (larger magnitude) chain weights $-J_{chain}$,  which could compress the logical problem scale in such a way that logical solution quality degrades relative to physical  solution quality.  However, this proposed mechanism does not correlate perfectly with observedperformance in general;  more work is needed to fully characterize and understand this phenomenon.  

Finally, we note that this unembedding penalty is not necessarily prohibitively damaging to logical solution quality.  The bottom SK panel shows a clear 
horizontal point stratification that 
correspond to the spectrum of optimal and nearest-optimal solution energies for this input.  Although the QPU did not find ground states in this particular test,  it found several solutions within fifth- or sixth-best overall.  Relative performance 
on SK inputs may be more due to classical ``easiness''  due to properties of small cliques, than of quantum ``hardness''  due to long chains.   

\subsection{Test Scope}
\label{scope} 

The specific numerical results reported in Sections \ref{results} and \ref{discussion} should not be considered 
definitive, since they depend on the autotuning policies applied under 
our Fair Test procedure (Section 
\ref{solvers}).  Our study, which focuses on robust performance over a broad variety of input types,  
considers parameter tuning to be type of computational overhead that our test design attempts to minimize.  
Development of benchmark tests that consider quantum and classical solver efficiency on specific problem domains,  under comparable levels of tuning effort, would be an interesting direction for future research. 

We believe, however,  that the observed general patterns of relative performance and convergence, 
and our discussion of mechanisms driving those patterns,  can be extended to other general-purpose heuristic methods that similarly do not exploit input-specific assumptions.  We believe this category of heuristics is the most appropriate comparison group for our envisioned use case, which requires robust performance from general-purpose BQM solvers deployed over the cloud.  

As discussed in Section \ref{inputs}, some input classes were not included in our main tests, because instances small enough to be embeddable on the Advantage QPU were too small (and therefore too easy) for tests of Milestone 2 to 
be viable.  Inputs from other application domains, such as problems with global constraints, which carry higher 
translation and embedding overhead costs than those studied here, could not be included for similar reasons.  
Although such ``too easy'' inputs will always exist, we believe they are not 
likely to arise in real-world applications, 
and {\em ipso facto} not interesting candidates for demonstrations of quantum utility.  

As mentioned in the introduction, we consider Milestones 1 and 2 to be most important for demonstrating quantum  utility because the other overhead times in Figure \ref{fig::layers} may be considered negligible or irrelevant, depending on the intended use case.  For example, the direct cost of minor embedding can sometimes 
be amortized over many runs on same-structured graphs, such as: cliques, used in wireless network applications \cite{kim20};  regular lattice-like structures, used in materials simulation \cite{king22};  circuits, used in fault detection \cite{bianfault, perdomofault};  and maps, used in routing problems \cite{clark-robots, hari-cvrp}.  

For another example, the overhead costs of network transmission may apply equally to both quantum and 
classical solution methods,  and use cases may prioritize either network latency or throughput.  For yet another, 
commercial users may prioritize performance metrics unrelated to computation time, such as the dollar cost of incorporating a given solution approach into an existing workflow.  

In this context, application scenarios for which Milestones 1 and 2 might be considered sufficient 
to demonstrate quantum utility would have the following properties: (a) they use large lattice-like inputs that 
can be directly mapped to the hardware graph or  else minor-embedded with short chains;  (b) they require many 
independent solution samples in short time frames; and (c) additional overheads in Figure \ref{fig::layers} are 
not important to the comparison.  As it turns out, these criteria have been met by some research applications in 
quantum materials simulation, where D-Wave systems have been shown to outperform classical simulation 
solvers by several orders of magnitude \cite{kingsim}.  

\section{Conclusions} 
\label{conclusions}

We introduce quantum utility, an approach to quantum performance evaluation on optimization problems, 
that aims to capture the user experience by explicitly considering overhead costs attached to 
the quantum computation.  We identify three milestones associated
with specific categories of overheads, and present experimental results for Milestones 1 and 2, 
which measure performance of an Advantage QPU against seven classical solvers running on 
CPUs and GPUs, using a testbed of 13 input classes.  

Using a rank-based performance metric that identifies solvers outcomes as  ``wins'' or ``fails",  
we observe that the Advantage QPU wins in 99\% tests  of Milestone 1  and in 19\% of tests of Milestone 2. 
Cases where the QPU wins are generally the same cases where classical solvers most often fail, and are associated with 
largest problem sizes $n$ and smallest limits on time per solution $t/s$.   Section \ref{discussion} describes some fundamental distinctions between classical and quantum performance mechanisms that explain these performance differences.  

We believe that our understanding of classical and quantum performance drivers bodes well for 
continued progress towards the goal of demonstrating quantum utility on broad 
categories of problems using future-generation QPUs.  This belief is based on early experience with a small 
prototype (500 qubits) of the Advantage2 system,\footnote{The Advantage2 prototype is available to the 
public through D-Wave's Leap portal.  Full sized Advantage2 systems are expected to launch in 2023 or 2024.} which suggests that full-sized QPUs (7000 qubits) will hold larger inputs with more compact embeddings (shorter chains), and will demonstrate technological improvements (better noise suppression and larger energy scale) yielding lower effective temperatures\cite{lowernoisewp, zephyrtr, adv2tr}.  

Furthermore, access time overheads, which have slightly decreased over the past decade as qubit counts 
have grown from 500+  on the D-Wave Two \cite{mcgwan}, to 5000+  on the Advantage processor, are not expected to increase significantly in the foreseeable future.  
 
Thus, replicating our Milestone 2 tests on the same and larger inputs, under identical time constraints, should expose degraded classical performance (from traversing smaller proportions of larger search spaces), and improved QPU performance (from lower effective temperatures and shorter chains), thereby increasing the variety of inputs and test scenarios that  
see QPU wins and classical fails.  As well, new types of input will pass 
our hardness test (Section \ref{inputs}) and qualify to enter the benchmarking arena.   We look forward to developing new benchmark tests 
incorporating  broader
varieties of inputs and more challenging milestones using future generations of annealing-based quantum systems.


\bibliographystyle{plain}
\bibliography{ccm_utility} 




\end{document}